\documentstyle[12pt,epsf]{article}

\def\Journal#1#2#3#4{{#1} {\bf #2}, #3 (#4)}

\def\PLB{{\em Phys. Lett.}  B}
\def\PRL{\em Phys. Rev. Lett.}
\def\PRD{{\em Phys. Rev.} D}

\def\ra{\rightarrow}

\def\al{\alpha}

\def\be{\begin{equation}}
\def\ee{\end{equation}}
\def\bea{\begin{eqnarray}}
\def\eea{\end{eqnarray}}

\begin{document}
\topskip 2cm 

\hspace*{\fill}\parbox[t]{4cm}{EDINBURGH 98/7\\ May 1998} 

\vspace{2cm}

\begin{center}
{\large\bf QCD Amplitudes in the High-Energy Limit} \\
\vspace{1.5cm}
{\large Vittorio Del Duca}\footnote{On leave of absence from
I.N.F.N., Sezione di Torino, Italy.} \\
\vspace{.5cm}
{\sl Particle Physics Theory Group,\,
Dept. of Physics and Astronomy\\ University of Edinburgh,\,
Edinburgh EH9 3JZ, Scotland, UK}\\

\vspace{1.5cm}

\begin{abstract}

Dijet production with a rapidity gap between the jets is
considered as a test ground for the production of a heavy Higgs
boson via weak-boson fusion at hadron supercolliders.
It is argued that in order to perform a detailed analysis of dijet 
production with a rapidity gap we need an ${\cal O}(\al_s^4)$
calculation including the relevant collinear enhancements
which give structure to the jets. Such a calculation
needs not be exact, but must include the full leading power in 
$\hat{s}/\hat{t}$ of the rate of dijet production in the high-energy
limit. The QCD amplitudes must be determined to the
corresponding accuracy. Accordingly, the
scattering amplitudes necessary to compute the full leading power in 
$\hat{s}/\hat{t}$ of dijet production to ${\cal O}(\al_s^3)$
are analysed.

\end{abstract}
\end{center}

\section{Rapidity gaps between jets}
\label{sec:0}

In recent years strong-interaction processes characterised by two
large and disparate energy scales, which are tipically the squared
center-of-mass energy $s$ and the squared momentum transfer $t$,
with $s\gg t$, have been extensively analysed. The interest stems
from the presumption
that their description in terms of perturbative-QCD calculations
at a fixed order in the coupling constant $\al_s$ might not be
adequate, and that a resummation at all orders of $\al_s$ of large 
contributions of the type of $\ln(s/t)$, performed through the
BFKL equation \cite{fkl}, might be needed.

These processes can be divided in two categories: $a)$ inclusive
processes, like deeply inelastic scattering (DIS) at small $x_{bj}$,
dijet production in $p\bar p$ collisions at large rapidity intervals,
forward jet production in DIS; $b)$ diffractive processes, like
diffractive DIS, diffractive vector meson production, or dijet production in 
hadron collisions with a rapidity gap between the tagging jets.

The last of these processes, dijet production with a rapidity gap,
is an example of double hard diffraction and is characterised by two
large and disparate energy scales, the squared parton
center-of-mass energy $\hat s$ and a momentum-transfer scale of the
order of the squared transverse energy of the jets $E^2_\perp$,
and by a soft scale, $\mu_s$, the threshold energy for the detection of
secondary hadrons within the rapidity gap. This process has been
studied both at the Tevatron Collider \cite{tev} and in
photoproduction at HERA \cite{zeus}. Since the formation of a
rapidity gap can happen only through the exchange of a colour singlet,
which could be modeled by the BFKL resummation \cite{mt,dt}, it is obvious
in that respect the interest for dijet production with a rapidity gap.

However, the main motivation for the analysis of
dijet production with a rapidity gap is to use it as a test
ground for the production of a heavy Higgs boson at hadron supercolliders
\cite{bj}. A Higgs boson is mainly
produced via gluon fusion, $g\,g\ra H$, mediated by a top-quark loop.
The Higgs boson then decays mainly into a pair of $W$
or $Z$ bosons. Such a signal, though, is going to be swamped by the $W\,W$
QCD and the $t\,\bar t$ backgrounds. A heavy Higgs boson is also
produced via weak-boson fusion, $W\,W,\; Z\,Z\ra H$, though at a
smaller rate~\cite{gun}, however such a production mechanism would have a
distinctive radiation pattern with a gap in parton production in the
central-rapidity region, because no color is exchanged between the 
quarks that emit the weak bosons \cite{bj,dkt}.

Producing a rapidity gap at the parton level is not sufficient though,
since the gap is usually filled by soft hadrons produced in the rescattering
between the spectators partons in the underlying event. Accordingly,
Bjorken~\cite{bj} introduced the gap survival probability, $<|S^2|>$,
i.e. the probability for a gap formed at the parton level to survive
the rescattering between the spectators partons. The gap survival 
probability deals with the soft, i.e. low transverse-momentum,
physics of the scattering between the two hadrons; therefore it can
only be estimated using non-perturbative models \cite{bj,glm}
(within perturbative QCD, the necessity of fulfilling the
factorization theorems \cite{css} would always allow for the emission
of soft hadrons in the rescattering between the spectators partons \cite{dt}).

In addition, at the CERN LHC collider the
requirement of running at very high luminosity will yield
overlapping events in the same bunch crossing which are an additional
source of soft hadrons and will further, and hopelessly, suppress the 
gap signal. A way out of this deadlock is to require a gap in minijet 
production~\cite{dks,bpz} rather than in soft-hadron production.
This has the further advantage of dispensing with the
gap survival probability, because the production of soft hadrons 
in the rescattering between the spectator partons is unrestricted,
since the transverse energy of the minijets is of $O$(10 GeV).

Modelling dijet production with a rapidity gap at the parton level
is not easy: a rapidity gap in parton-parton scattering could
be produced via $\gamma,\,W,\,Z$-boson exchange in the crossed channel;
however, their rates turn out to be too small~\cite{che}. Single
gluon exchange in the crossed channel, on the other hand, 
is likely not to produce a gap because the exchanged 
gluon being a color octet radiates off more gluons. 
Exchanging two gluons in the crossed channel in a color-singlet 
configuration is the simplest way of producing a gap at the parton
level. However, this is an ${\cal O}(\al_s^4)$ process, for which no
detailed calculation is available yet.

Bjorken~\cite{bj} made an estimate of the order of magnitude of the 
rate for two-gluon exchange to be about 10\% of the 
one-gluon exchange rate, $\hat\sigma_{sing}/\hat\sigma_{oct} \sim 0.1\,$.
In the limit of $\hat{s}\gg \hat{t}$, Chehime and
Zeppenfeld \cite{chez} made an $O(\al_s^5)$ analysis of the bremsstrahlung
pattern of a gluon emitted in parton-parton scattering with two-gluon
singlet exchange; they found 
that if the transverse momentum $p_{\perp rad}$ of the 
radiated gluon is of the same order as the transverse momenta $p_{\perp jet}$
of the tagging jets then the gluon is radiated mainly in the 
central-rapidity region, like in the one-gluon exchange case; if
$p_{\perp rad} \ll p_{\perp jet}$, then the gluon is radiated mainly in the
forward direction; this is in agreement with the classical expectation
that if the bremsstrahlung
gluon is hard it has a short wavelength and may resolve the color structure
of the two-gluon exchange; if it is soft its resolving power is low and
sees the two exchanged gluons as a color singlet.

In addition, in the limit of $\hat{s}\gg \hat{t}$, the leading
logarithmic (LL) contributions in $\ln(\hat{s}/|\hat{t}|)$ to the radiative
corrections to dijet production with a rapidity gap can be
resummed at all orders of $\al_s$ through the BFKL equation.
Resumming the leading virtual radiative corrections
to one-gluon exchange in parton-parton scattering, 
one obtains a Sudakov suppression~\cite{dt};
namely, the production rate decreases exponentially as the rapidity
gap width $\Delta y\simeq \ln(\hat{s}/|\hat{t}|)$ increases. 
On the contrary, the resummation of the leading virtual radiative corrections
to two-gluon singlet exchange yields a production rate~\cite{mt,dt}
\be
{d\hat\sigma_{sing}\over d\hat{t}} \simeq 
{81\pi^3\al_s^4\over 4\hat{t}^2}\, {\exp\left[24\ln{2}\al_s\, 
\Delta y/\pi\right]\over \left[21\zeta(3)\al_s
\Delta y/2\right]^3}\, ,\label{dinov}
\ee
with $\zeta(3) = 1.20206...$, which is initially approximately flat,
due to the competing effects of $\Delta y$ in the exponential in the
numerator and in the cube in the denominator, then it
increases exponentially with $\Delta y$.
This is in qualitative agreement with
the data~\cite{tev}, which show that the rate of 
dijet production with a rapidity gap falls off exponentially
at small gap widths (when the scattering is dominated by one-gluon
exchange), and it flattens out at larger gap widths (when the
scattering starts being dominated by two-gluon singlet exchange).
However, no final rise of the gap production rate is observed.

The BFKL resummation, used in ref.~\cite{mt,dt}, as well the analysis
of ref.~\cite{chez} in the high-energy limit $\hat{s}\gg \hat{t}$,
are not suitable for a detailed study of jet production because they
are leading order analyses and their jets are leading partons.
This drawback is even more acute for dijet production with a rapidity gap
because the experiments measure the pseudorapidity gap width, 
$\Delta\eta_c$, between the edges of the jet cones,
which differs from the pseudorapidity difference, $\Delta\eta$, between the 
jet centers by the cone sizes $R$, $\Delta\eta_c = \Delta\eta - 2R$. 
The BFKL approximation is not able to distinguish between $\Delta\eta$ 
and $\Delta\eta_c$. 

In order to examine the gap fraction as a function of the gap width between
the jet-cone edges, while accounting properly for the cone structures, we
need a next-to-leading order calculation which includes, though, the basic 
features of color-singlet exchange. As we have mentioned, the simplest
calculation of this kind 
for the dijet production rate with a rapidity gap is ${\cal O}(\al_s^4)$.
At the moment an {\em exact} calculation of this kind is unfeasible 
because one needs to know two-loop parton-parton scattering amplitudes, 
which have not been computed yet.
However, we want to argue that in order to analyse the basic 
features of color-singlet exchange it is not necessary to have an exact
${\cal O}(\al_s^4)$ calculation; instead, it should be enough to know
the full leading power in $\hat{s}/\hat{t}$ of the rate of dijet production
with a rapidity gap. This will contain all the
collinear enhancements which give structure to the jets.
Such calculation is not available yet either;
however, albeit daunting, it should be much simpler than an exact
${\cal O}(\al_s^4)$ calculation of dijet production.

In the next section we shall show that all the relevant collinear enhancements
which appear in the exact ${\cal O}(\al_s^3)$ calculation of dijet 
production are already present in the next-to-leading-logarithmic (NLL)
corrections to the parton-parton scattering amplitudes.

\section{QCD amplitudes in the high-energy limit}
\label{sec:1}

Let us consider the BFKL resummation \cite{fkl}, i.e. the 
LL approximation in $\ln(\hat{s}/|\hat{t}|)$, applied
to dijet production; the QCD amplitudes we need in order to compute 
the BFKL resummation are of two kinds: $a)$ virtual corrections to
parton-parton scattering, in the high-energy limit $\hat{s}\gg\hat{t}$
and in LL approximation; $b)$ real corrections to
parton-parton scattering, in the multi-Regge kinematics, i.e. in the
strong rapidity ordering of the produced partons; when the scattering
amplitudes are integrated over the phase space with multi-Regge 
kinematics, they yield the right power of $\ln(\hat{s}/|\hat{t}|)$ 
to match the virtual corrections in LL approximation. 
The strong rapidity 
ordering constrains the scattering amplitudes not to have collinear
singularities, because the squared invariant mass, $m_{ij}^2$, of 
any two produced partons is constrained to be large $m_{ij}^2 \simeq 
|p_{i\perp}| |p_{j\perp}| \exp(|y_i-y_j|)$; accordingly, the
LL approximation lacks the running of the coupling constant,
which must be considered fixed. Infrared singularities, though,
are present in the amplitudes, both real and virtual, but they cancel
after these are put together to compute the radiative corrections to
dijet production.

\subsection{QCD amplitudes in multi-Regge kinematics}
\label{sec:1a}

A tree-level multigluon amplitude in a helicity basis can be written
as a sum over color-ordered permutations of a color factor times a
subamplitude, which is dependent on the helicities and momenta of the
external partons \cite{mp}
\begin{equation}
M_n = \sum_{[a,1,...,n,b]'} {\rm tr}(\lambda^a\lambda^{d_1} \cdots
\lambda^{d_n} \lambda^b) \, m(-p_a,-\nu_a; p_1,\nu_1;...;
p_n,\nu_n; -p_b,-\nu_b)\, ,\label{one}
\end{equation}
where $a,d_1,..., d_n,b$, and $\nu_a,\nu_1,...,\nu_b$
are respectively the colors and the helicities of the gluons,
the $\lambda$'s are the color matrices in the fundamental representation
of SU($\rm N_c$), the sum is over the noncyclic permutations of the 
color orderings $[a,1,...,b]$ and all the momenta are taken as outgoing. 
In particular, for the {\sl maximally helicity-violating}
configurations $(-,-,+,...,+)$, the subamplitudes
$m(-p_a,-\nu_a; p_1,\nu_1;...; p_n,\nu_n; -p_b,-\nu_b)$, 
assume the form \cite{pt},
\begin {equation}
m(-,-,+,...,+) = 2^{1+n/2}\, g^n\, {\langle p_i p_j\rangle^4\over
\langle p_a p_1\rangle \cdots\langle p_n p_b\rangle 
\langle p_b p_a\rangle}\, ,\label{two}
\end{equation}
with $i$ and $j$ the gluons of negative helicity. 

Before considering multiparton amplitudes in multi-Regge kinematics,
we need to recall how a parton-parton scattering amplitude factorises
in the high-energy limit.
We consider the elastic scattering of two gluons of momenta $p_a$ and $p_b$
in two gluons of momenta $p_{a'}$ and $p_{b'}$, in the limit
$\hat s\gg |\hat t|$. The corresponding amplitude is (cf. ref.~\cite{thuile} 
and the Appendix), 
\be
M^{aa'bb'}_{\nu_a\nu_{a'}\nu_{b'}\nu_b} = 2 {\hat s}
\left[i g\, f^{aa'c}\, C^{gg}_{-\nu_a\nu_{a'}}(-p_a,p_{a'}) \right]
{1\over \hat t} \left[i g\, f^{bb'c}\, C^{gg}_{-\nu_b\nu_{b'}}(-p_b,p_{b'}) 
\right]\, ,\label{elas}
\ee
with $p_{b\perp} = -p_{a\perp} = q_\perp$ and $\hat t \simeq -|q_\perp|^2$,
and the helicity-conserving vertices $g^*\, g \rightarrow g$, 
with $g^*$ an off-shell gluon,
\be
C_{-+}^{gg}(-p_a,p_{a'}) = 1 \qquad C_{-+}^{gg}(-p_b,p_{b'}) =
{p_{b'\perp}^* \over p_{b'\perp}}\, .\label{centrc}
\ee
The $C$-vertices transform
into their complex conjugates under helicity reversal,
$C_{\{\nu\}}^*(\{k\}) = C_{\{-\nu\}}(\{k\})$. The helicity-flip
vertex $C_{++}$ is subleading in the high-energy limit.

The quark-gluon $q\, g \rightarrow q\, g$
scattering amplitude in the high-energy limit is,
\bea
M^{q\, g \rightarrow q\, g} &=& 2 {\hat s} \left[g\, \lambda^c_{a' \bar a}\,
C_{-\nu_a \nu_a}^{\bar q q}(-p_a,p_{a'}) \right] {1\over \hat t} 
\left[i g\, f^{bb'c}\, C^{gg}_{-\nu_b\nu_{b'}}(-p_b,p_{b'}) \right]\, 
,\label{elasqa} \\
M^{g\, q \rightarrow g\, q} &=& 2 {\hat s} \left[i g\, f^{aa'c}\, 
C^{gg}_{-\nu_a\nu_{a'}}(-p_a,p_{a'}) \right] {1\over \hat t}
\left[g\, \lambda^c_{b' \bar b}\,
C_{-\nu_b \nu_b}^{\bar q q}(-p_b,p_{b'}) \right]\, ,\label{elasqb}
\eea
where we have labelled the incoming quarks
as outgoing antiquarks by convention, and
the antiquark is $-p_a$ in eq.(\ref{elasqa}) 
and $-p_b$ in eq.(\ref{elasqb}),
and the $C$-vertices $g^*\, q \rightarrow q$ are,
\be
C_{-+}^{\bar q q}(-p_a,p_{a'}) = 1\, ;\qquad C_{-+}^{\bar q q}(-p_b,p_{b'}) =
\left({p_{b'\perp}^* \over p_{b'\perp}}\right)^{1/2}\, .\label{cbqqm}
\ee
The antiquark-gluon $\bar q\, g \rightarrow \bar q\, g$
amplitude is found accordingly \cite{thuile}. 
The quark-quark $q\,q\to q\,q$ scattering amplitude in the high-energy
limit is
\begin{equation}
M^{q\,q\to q\,q} = 2 s \left[g\, 
\lambda^c_{a' \bar a}\,
C_{-\nu_a \nu_a}^{\bar q q}(-p_a,p_{a'}) \right] {1\over t} 
\left[g\, \lambda^c_{b' \bar b}\,
C_{-\nu_b \nu_b}^{\bar q q}(-p_b,p_{b'}) \right]
.\label{elasq}
\end{equation}
The amplitudes (\ref{elas}), (\ref{elasqa}), (\ref{elasqb}), 
(\ref{elasq}), have all the
effective form of a gluon exchange in the $t$ channel,
and differ only for the relative color strength in the production vertices
\cite{CM}.
This allows us to replace an incoming gluon with a quark, for instance
on the upper line, via the simple substitution
\be
i g\, f^{aa'c}\, C^{gg}_{-\nu_a\nu_{a'}}(-p_a,p_{a'}) \leftrightarrow g\, 
\lambda^c_{a' \bar a}\, C_{-\nu_a \nu_a}^{\bar q q}(-p_a,p_{a'})\, 
,\label{qlrag}
\ee
and similar ones for an antiquark and/or for the lower line.

Next, we consider the production of three gluons of momenta $p_{a'}$,
$k$ and $p_{b'}$, and we require that the gluons are 
strongly ordered in their rapidities and have comparable transverse momenta,
\begin{equation}
y_{a'} \gg y \gg y_{b'};\qquad |p_{a'\perp}|\simeq|k_\perp|\simeq
|p_{b'\perp}|\, .\label{treg}
\end{equation}
Eq.(\ref{treg}) is the simplest example of {\sl multi-Regge kinematics}
(\ref{mreg}). The amplitude is,
\begin{eqnarray}
& & M^{g g \ra g g g}(-p_a,-\nu_a; p_{a'},\nu_{a'}; k,\nu; p_{b'},
\nu_{b'}; -p_b,-\nu_b) \nonumber\\ 
&=& 2 {\hat s} \left[i g\, f^{aa'c}\, C_{-\nu_a\nu_{a'}}^{gg}
(-p_a,p_{a'}) \right]\, {1\over\hat t_1} \label{three}\\ &\times& 
\left[i g\, f^{cdc'}\, C^g_{\nu}(q_1,q_2)\right]\, {1\over \hat t_2}\, 
\left[i g\, f^{bb'c'}\, C_{-\nu_b\nu_{b'}}^{gg}(-p_b,p_{b'}) \right]\,
,\nonumber
\end{eqnarray}
with $p_{a'\perp} = - q_{1\perp}$, $p_{b'\perp} = q_{2\perp}$ and
$\hat t_i \simeq - |q_{i\perp}|^2$ with $i=1,2$ and with Lipatov vertex
$g^*\, g^* \rightarrow g$ \cite{lipat,ptlip},
\be
C^g_+(q_1,q_2) = \sqrt{2}\, {q^*_{1\perp} q_{2\perp}\over k_\perp}\, 
.\label{lip}
\ee
The amplitude (\ref{three}) has the effective form of a gluon-ladder 
exchange in the $t$ channel.
Again, we may replace an incoming gluon with a quark
via the substitution (\ref{qlrag}). As we shall see in 
sect.~\ref{sec:1b}, no quarks may be produced along the ladder
since that would involve quark exchange in the $t$ channel, which is
suppressed in the kinematics (\ref{treg}).
Eq.(\ref{three}) generalizes to the production of $n$ gluons  
in multi-Regge kinematics (\ref{mreg}) in a straightforward manner
\cite{fkl,ptlip}.

In the soft limit, $k \ra 0$, a generic subamplitude in eq.~(\ref{one}) 
factorises as \cite{mp,bg}, 
\bea
& &\lim_{k\ra 0} m^{g g \ra g g g}(-p_a,-\nu_a; p_{a'},\nu_{a'}; k,\nu; 
p_{b'},\nu_{b'}; -p_b,-\nu_b) \nonumber\\
& & = m^{g g \ra g g}(-p_a,-\nu_a; p_{a'},\nu_{a'};
p_{b'},\nu_{b'}; -p_b,-\nu_b) \times {\rm soft}(p_{a'}; k,\nu; 
p_{b'})\, ,\label{soft}
\eea
with eikonal factor \cite{mp,bcm},
\be
{\rm soft}(p_{a'}; k,+; p_{b'}) = \sqrt{2}\, 
{\langle p_{a'} p_{b'}\rangle \over \langle p_{a'} k\rangle 
\langle k p_{b'}\rangle}\, ,\label{sfac}
\ee
with spinor products (\ref{spro}), and
${\rm soft}(p_{a'}; k,-; p_{b'})$ obtained from ${\rm soft}(p_{a'}; 
k,+; p_{b'})$ by exchanging the $\langle j k\rangle$ spinor products
in eq.~(\ref{sfac}) with $[k j]$ products. The eikonal factor (\ref{sfac})
does not depend on the helicities of gluons $a'$ and $b'$. Note that
the soft factorization of eq.~(\ref{soft}) does not carry on the
full amplitude (\ref{one}). Indeed, for the helicity configurations
allowed by eq.~(\ref{soft}), eq.~(\ref{three}) reduces to
\bea
& &\lim_{k\ra 0} M^{g g \ra g g g}(-p_a,-\nu_a; p_{a'},\nu_{a'}; k,\nu; 
p_{b'},\nu_{b'}; -p_b,-\nu_b) \nonumber\\
&=& 2 {\hat s} \left[i g\, f^{aa'c}\, C_{-\nu_a\nu_{a'}}^{gg}
(-p_a,p_{a'}) \right]\, {1\over\hat t} \label{sthree}\\ &\times& 
\left[ i g\, f^{cdc'}\, {\rm soft}(p_{a'}; k,\nu; 
p_{b'}) \right]\,
\left[i g\, f^{bb'c'}\, C_{-\nu_b\nu_{b'}}^{gg}(-p_b,p_{b'}) \right]\,
,\nonumber
\end{eqnarray}
with
\be
{\rm soft}(p_{a'}; k,+; p_{b'}) = - {\sqrt{2}\over k_\perp}\, ,\label{mrsfac}
\ee
and ${\rm soft}(p_{a'}; k,-; p_{b'}) = {\rm soft}(p_{a'}; k,+; p_{b'})^*$.

The square of the amplitude (\ref{three}), 
integrated over the phase space of the intermediate gluon in
multi-Regge kinematics (\ref{treg}) yields an 
$O(\alpha_s\ln( s/| t|))$ correction to gluon-gluon scattering.
Because of eq.~(\ref{soft}), the real correction is infrared divergent.
To complete the $O(\alpha_s)$ corrections, and to cancel the infrared
divergence, one needs the 1-loop gluon-gluon amplitude
in LL approximation.
The virtual radiative corrections to eq.~(\ref{elas}) in
LL approximation are obtained, to all orders
in $\alpha_s$, by replacing \cite{fkl},
\begin{equation}
{1\over t} \to {1\over t} 
\left({s\over -t}\right)^{\alpha(t)}\, ,\label{sud}
\end{equation}
in eq.~(\ref{elas}), with $\alpha(t)$ related to the loop 
transverse-momentum integration
\begin{equation}
\alpha(t) = \alpha_s\, N_c\, \hat{t} \int 
{d^2k_{\perp}\over (2\pi)^2}\, {1\over k_{\perp}^2
(q-k)_{\perp}^2}\, ,\label{allv}
\end{equation}
and $\alpha_s = g^2/4\pi$. 
The infrared divergence in eq.~(\ref{allv}) can be regulated in 4
dimensions with an infrared-cutoff mass. Alternatively, 
using dimensional regularization
in $d=4-2\epsilon$ dimensions, the integral
in eq.~(\ref{allv}) is performed in $2-2\epsilon$ dimensions, yielding
\begin{equation}
\alpha(t) = 2 g^2\, N_c\, {1\over\epsilon} 
\left(\mu^2\over -t\right)^{\epsilon} c_{\Gamma}\, ,\label{alph}
\end{equation}
with
\begin{equation}
c_{\Gamma} = {1\over (4\pi)^{2-\epsilon}}\, {\Gamma(1+\epsilon)\,
\Gamma^2(1-\epsilon)\over \Gamma(1-2\epsilon)}\, .\label{cgam}
\end{equation}
Adding the 1-loop gluon-gluon 
amplitude, multiplied by its tree-level counterpart,
to the square of the amplitude (\ref{three}), 
integrated over the phase space of the intermediate gluon, cancels
the infrared divergences and yields a finite $O(\alpha_s\ln(
s/| t|))$ correction to gluon-gluon scattering.

\subsection{Real NLL corrections to QCD amplitudes}
\label{sec:1b}

In order to compute the NLL corrections 
to the BFKL resummation \cite{nll}, one needs: $a)$ to compute the
virtual corrections to NLL accuracy; $b)$ to relax the strong
rapidity ordering in the real corrections, by allowing any two
partons to be produced with comparable rapidities.
I shall illustrate how this comes about in
a fixed-order calculation, by considering the NLL ${\cal O}(\alpha_s)$
corrections to parton-parton scattering. Let three partons be
produced with momenta $k_1$, $k_2$ and $p_{b'}$ 
in the scattering between two partons of momenta $p_a$ and $p_b$, and
to be specific, I shall take partons $k_1$ and $k_2$ in the forward-rapidity
region of parton $p_a$, the analysis for $k_1$ and $k_2$ 
in the forward-rapidity region of $p_b$ being analogous,
\begin{equation}
y_1 \simeq y_2 \gg y_{b'}\,;\qquad |k_{1\perp}|\simeq|k_{2\perp}|
\simeq|p_{b'\perp}|\, .\label{qmreg}
\end{equation}
First we consider the amplitude for the scattering $g\, g\, \ra g\,
g\, g$, \cite{fl,ptlipnl}
\begin{eqnarray}
& & M^{gg}(-p_a,-\nu_a; k_1,\nu_1; k_2,\nu_2; p_{b'},\nu_{b'}; -p_b,-\nu_b)
\label{nllfg}\\ & & = 2\hat s \left\{ C^{g\,g\,g}_{-\nu_a\nu_1\nu_2}
(-p_a,k_1,k_2) \left[ (ig)^2\, f^{ad_1c} f^{cd_2c'} 
A_{\Sigma\nu_i}(-p_a,k_1,k_2,q) + \left(\begin{array}{c} k_1
\leftrightarrow k_2\\ d_1\leftrightarrow d_2 \end{array}\right)
\right] \right\} \nonumber\\ & & \times
{1\over \hat t}\, \left[ig\, f^{bb'c'} C^{g\,g}_{-\nu_b\nu_{b'}}(-p_b,p_{b'})
\right]\, ,\nonumber
\end{eqnarray}
with $C^{g\,g}$ as in eq.~(\ref{cbqqm}), and
the production vertex $g^*\, g \rightarrow g\, g$
of gluons $k_1$ and $k_2$ enclosed
in curly brackets, and with $\sum\nu_i=-\nu_a+\nu_1+\nu_2$ and
\begin{eqnarray}
C^{g\,g\,g}_{-++}(-p_a,k_1,k_2) = 1\, ; & & C^{g\,g\,g}_{+-+}(-p_a,k_1,k_2) 
= {1 \over\left(1+{k_2^+\over k_1^+}\right)^2}\, ;\nonumber\\ 
C^{g\,g\,g}_{++-}(-p_a,k_1,k_2) = {1 \over\left(1+{k_1^+\over k_2^+} 
\right)^2}\, ; & & A_+(-p_a,k_1,k_2,q) = - \sqrt{2}\, {q_\perp\over k_{1\perp}}
\sqrt{k_1^+\over k_2^+}\, {1\over \langle k_1 k_2\rangle}\, ;\label{ab}
\end{eqnarray}
where the spinor product $\langle k_1 k_2\rangle$ is defined in 
eq.(\ref{spro}), and the momentum exchanged in the
crossed channel is $q = p_{b'} - p_b$. The vertex
$C^{g\,g\,g}_{+++}(-p_a,k_1,k_2)$ 
is subleading in the high-energy limit.

In the multi-Regge limit $k_1^+\gg k_2^+$ 
the production vertex $g^*\, g \rightarrow g\, g$ becomes
\begin{eqnarray}
\lim_{k_1^+\gg k_2^+} C^{g\,g\,g}_{-\nu_a\nu_1\nu_2}
(-p_a,k_1,k_2) A_{\Sigma\nu_i}(-p_a,k_1,k_2,q) = 
C^{g\,g}_{-\nu_a\nu_1}(-p_a,k_1)
{1\over \hat t_1}\, C^g_{\nu_2}(q_1,q)\, ,\label{camr}
\end{eqnarray}
with $q_1 = p_a - k_1$, and $\hat t_1 \simeq - |q_{1\perp}|^2$,
thus the amplitude (\ref{nllfg}) reduces to the amplitude in
multi-Regge kinematics (\ref{three}), as expected. 

The collinear factorization for a 
generic amplitude occurs both on the subamplitude and on
the full amplitude \cite{mp}, since in eq.~(\ref{one}) color
orderings where the collinear gluons are not adjacent do not have 
a pole. Hence in the collinear limit for gluons
$i$ and $j$, with $k_i = z P$ and $k_j = (1-z) P$,
a generic amplitude (\ref{one}) can be written as
\be
\lim_{p_i || p_j} M^{... d_i d_j ...}(...; p_i,\nu_i; p_j,\nu_j; ...)
= M^{... c ...}(...; P,\nu; ...) {\rm Split}_{-\nu}(p_i,\nu_i; p_j,
\nu_j)\, .\label{gencoll}
\ee
Accordingly, in the collinear
limit, $k_1 = z P$ and $k_2 = (1-z) P$, using the Jacobi identity for
the group structure functions we can write the amplitude (\ref{nllfg}) as
\bea
& &\lim_{k_1 || k_2} M^{g g \ra g g g}(-p_a,-\nu_a; k_1,\nu_1; k_2,\nu_2;
p_{b'},\nu_{b'}; -p_b,-\nu_b) \nonumber\\ & & = M^{g g \ra g g}
(-p_a,-\nu_a; P,\nu; p_{b'},\nu_{b'}; -p_b,-\nu_b) \times 
{\rm Split}_{-\nu}^{g \ra g g}(k_1,\nu_1; k_2,\nu_2)\, ,\label{coll}
\eea
with $M^{g g \ra g g}$ given in eq.~(\ref{elas}), and
with collinear factor
\be
{\rm Split}_{-\nu}^{g \ra g g}(k_1,\nu_1; k_2,\nu_2) =  i g\,
f^{cd_1d_2}\, {\rm split}_{-\nu}^{g \ra g g}(k_1,\nu_1; k_2,\nu_2)\, 
,\label{colfac}
\ee
with splitting factors \cite{mp},
\bea
{\rm split}_-^{g \ra g g}(k_1,+; k_2,+) &=& \sqrt{2}\, {1\over \sqrt{z(1-z)}
\langle k_1 k_2\rangle}\, \nonumber\\
{\rm split}_+^{g \ra g g}(k_1,-; k_2,+) &=& \sqrt{2}\,{z^2\over \sqrt{z(1-z)}
\langle k_1 k_2\rangle}\, \label{split}\\
{\rm split}_+^{g \ra g g}(k_1,+; k_2,-) &=& \sqrt{2}\, 
{(1-z)^2\over \sqrt{z(1-z)}\langle k_1 k_2\rangle}\, ,\nonumber
\eea
and ${\rm split}_{\nu}^{g \ra g g}
(k_1,-\nu_1; k_2,-\nu_2)$ obtained from ${\rm split}_{-\nu}
^{g \ra g g}(k_1,\nu_1; k_2,\nu_2)$ by exchanging $\langle k_1
k_2\rangle$ with $[k_2 k_1]$. 

In the soft limit, $k_1\ra 0$, using the factorization of the
subamplitude (\ref{soft}), eq.~(\ref{nllfg}) reduces 
for the non-null subamplitudes to
\begin{eqnarray}
& & \lim_{k_1\ra 0} M^{g g \ra g g g}(-p_a,-\nu_a; k_1,\nu_1; k_2,\nu_2; 
p_{b'},\nu_{b'}; -p_b,-\nu_b) = 2\hat s\, (ig)^2\, 
\nonumber\\ & & \times \left[ f^{ad_1c} f^{cd_2c'} 
{\rm soft}(-p_a; k_1,\nu_1; k_2) + f^{ad_2c} f^{cd_1c'}
{\rm soft}(k_2; k_1,\nu_1; p_{b'}) \right] \nonumber\\ & & \times
{1\over \hat t}\, \left[ig\, f^{bb'c'} C^{g\,g}_{-\nu_b\nu_{b'}}(-p_b,p_{b'})
\right]\, ,\label{nonsoft}
\end{eqnarray}
with eikonal factors,
\begin{eqnarray}
{\rm soft}(-p_a; k_1,+; k_2) &=&  \sqrt{2}\,
{\langle p_a k_2\rangle \over \langle p_a k_1\rangle 
\langle k_1 k_2\rangle} = \sqrt{2}\, {k_{2\perp}\over k_{1\perp}}
\sqrt{k_1^+\over k_2^+}\, {1\over \langle k_1 k_2\rangle}
\label{nonfac}\\ {\rm soft}(k_2; k_1,+; p_{b'}) &=& \sqrt{2}\, 
{\langle k_2 p_{b'}\rangle \over \langle k_2 k_1\rangle 
\langle k_1 p_{b'}\rangle} = \sqrt{2}\, \sqrt{k_2^+\over k_1^+}\, 
{1\over \langle k_2 k_1\rangle}\, ,\nonumber
\end{eqnarray}
and the eikonal factors for $(k_1,-)$ obtained from eq.~(\ref{nonfac})
by complex conjugation.

The amplitude for the production of a $q\bar q$ pair in the forward-rapidity
region of gluon $p_a$ is \cite{ptlipqq},
\begin{eqnarray}
& & M^{g\, g\ra \bar{q}q\, g}(-p_a,-\nu_a; k_1,\nu_1; k_2,-\nu_1; 
p_{b'},\nu_{b'}; 
-p_b,-\nu_b) \label{forwqq}\\ & & = 2\hat s \left\{g^2\, 
C^{g\,\bar{q}\,q}_{-\nu_a\nu_1-\nu_1}(-p_a,k_1,k_2) \left[\left(\lambda^{c'} 
\lambda^a\right)_{d_2\bar{d_1}} A_{-\nu_a}(k_1,k_2) + \left(\lambda^a 
\lambda^{c'}\right)_{d_2\bar{d_1}} A_{-\nu_a}(k_2,k_1) \right] \right\} 
\nonumber\\ & & \times {1\over \hat t}\, \left[ig\, f^{bb'c'} 
C^{g\,g}_{-\nu_b\nu_{b'}}(-p_b,p_{b'}) \right]\, ,\nonumber
\end{eqnarray}
with $k_1$ the antiquark, the production vertex $g^*\, g 
\rightarrow \bar q q$ in curly brackets, $A$ defined in eq.(\ref{ab}),
and $C^{g\,\bar{q}\,q}$ given by,
\begin{eqnarray}
C^{g\,\bar{q}\,q}_{++-}(-p_a,k_1,k_2) &=& \sqrt{k_1^+\over k_2^+}
{1 \over\left(1+{k_1^+\over k_2^+} \right)^2} \label{cqqa}\\
C^{g\,\bar{q}\,q}_{+-+}(-p_a,k_1,k_2) &=& \sqrt{k_2^+\over k_1^+}
{1 \over\left(1+{k_2^+\over k_1^+} \right)^2}\, .\nonumber
\end{eqnarray}

In the multi-Regge limit $k_1^+\gg k_2^+$ the vertex $g^*\, g 
\rightarrow \bar q q$ in eq.~(\ref{forwqq}) vanishes, i.e. as 
anticipated in sect.~\ref{sec:1a} quark production 
along the ladder is suppressed in the multi-Regge kinematics.

Like the purely gluonic amplitude (\ref{one}), also a generic
amplitude with a $\bar q q$ pair factorises in the collinear limit
according to eq.~(\ref{gencoll}). Accordingly, for
$k_1 = z P$ and $k_2 = (1-z) P$, eq.~(\ref{forwqq}) reduces to
\begin{eqnarray}
& & \lim_{k_1 || k_2} M^{g\, g\ra \bar{q}q\, g}(-p_a,-\nu_a; k_1,\nu_1; 
k_2,-\nu_1; p_{b'},\nu_{b'}; -p_b,-\nu_b) \label{colqq}\\ & & = 
M^{g g \ra g g}(-p_a,-\nu_a; P,\nu; p_{b'},\nu_{b'}; -p_b,-\nu_b) \times 
{\rm Split}_{-\nu}^{g \ra \bar{q}q}(k_1,\nu_1; k_2,-\nu_1)\, ,\nonumber
\end{eqnarray}
with $M^{g g \ra g g}$ given in eq.~(\ref{elas}), and with collinear factor
\be
{\rm Split}_{-\nu}^{g \ra \bar{q}q}(k_1,\nu_1; k_2,-\nu_1) =  g\,
(\lambda^c)_{d_2\bar{d_1}}\,
{\rm split}_{-\nu}^{g \ra \bar{q}q}(k_1,\nu_1; k_2,-\nu_1)\, .\label{qqfac}
\ee
with splitting factors \cite{mp},
\begin{eqnarray}
{\rm split}_+^{g \ra \bar{q}q}(k_1,+; k_2,-) &=& \sqrt{2}\, {z^{1/2}(1-z)^{3/2}
\over \sqrt{z(1-z)}\langle k_1 k_2\rangle} \nonumber\\
{\rm split}_+^{g \ra \bar{q}q}(k_1,-; k_2,+) &=& \sqrt{2}\, {z^{3/2}(1-z)^{1/2}
\over \sqrt{z(1-z)}\langle k_1 k_2\rangle}\, ,\label{qqsplit}
\eea
and with ${\rm split}_{\nu}^{g \ra \bar{q}q}
(k_1,-\nu_1; k_2,\nu_1)$ obtained from ${\rm split}_{-\nu}
^{g \ra \bar{q}q}(k_1,\nu_1; k_2,-\nu_1)$ by exchanging $\langle k_1
k_2\rangle$ with $[k_2 k_1]$. 

In the soft limits, $k_1 \ra 0$ or $k_2 \ra 0$, the vertex $g^*\, g 
\rightarrow \bar q q$ in eq.~(\ref{forwqq}) yields at most a 
square-root singularity, whose square once integrated over the
phase space does not yield an infrared divergence, in accordance to
the general feature that there are no infrared divergences 
associated to soft fermions.

The amplitude for the production of a $q\, g$ pair in the forward-rapidity
region of quark $p_a$ is 
\begin{eqnarray}
& & M^{q\,g\ra q\,g\,g}(-p_a,-\nu; k_1,\nu; k_2,\nu_2; p_{b'},\nu_{b'}; 
-p_b,-\nu_b) \label{forwqg}\\ & & = 2\hat s \left\{ g^2\, 
C^{\bar{q}\,q\,g}_{-\nu\nu\nu_2}(-p_a,k_1,k_2) \left[\left(\lambda^{d_2}
\lambda^{c'}\right)_{d_1\bar{a}} A_{\nu_2}(k_1,k_2) - \left(\lambda^{c'}
\lambda^{d_2}\right)_{d_1\bar{a}} B_{\nu_2}(k_1,k_2)\right] \right\} 
\nonumber\\ & & \times {1\over \hat t}\, \left[ig\, f^{bb'c'} 
C^{g\,g}_{-\nu_b\nu_{b'}}(-p_b,p_{b'}) \right]\, ,\nonumber
\end{eqnarray}
with $k_1$ the final-state quark, and the production vertex 
$q\, g^* \rightarrow q\, g$ in curly brackets, with $A$ defined in 
eq.(\ref{ab}), and $B$ given by
\be
B_{\Sigma\nu_i}(-p_a,k_1,k_2) = A_{\Sigma\nu_i}(-p_a,k_1,k_2) + 
A_{\Sigma\nu_i}(-p_a,k_2,k_1)\, ,\label{b}
\ee
and 
\begin{eqnarray}
C^{\bar{q}\,q\,g}_{-++}(-p_a,k_1,k_2) &=&
{1 \over\left(1+{k_2^+\over k_1^+} \right)^{1/2}}\, ,\label{cqg}\\
C^{\bar{q}\,q\,g}_{+-+}(-p_a,k_1,k_2) &=&
{1 \over\left(1+{k_2^+\over k_1^+} \right)^{3/2}}\, .\nonumber
\end{eqnarray}

In the multi-Regge limit $k_1^+\gg k_2^+$ the amplitude (\ref{forwqg})
reduces to eq.(\ref{three}), with the substitution 
(\ref{qlrag}) for the upper line, and the vertex 
$C^{\bar{q}\,q}$ in eq.(\ref{cbqqm}).

In the collinear limit, $k_1 = z P$ and $k_2 = (1-z) P$, the
coefficient $B$, eq.~(\ref{b}), vanishes and eq.~(\ref{forwqg}) 
reduces to
\begin{eqnarray}
& & \lim_{k_1 || k_2} M^{q\,g\ra q\,g\,g}(-p_a,-\nu; k_1,\nu; 
k_2,\nu_2; p_{b'},\nu_{b'}; -p_b,-\nu_b) \label{colqg}\\ & & = 
M^{q\, g \ra q\, g}(-p_a,-\nu; P,\nu; p_{b'},\nu_{b'}; -p_b,-\nu_b) \times 
{\rm Split}_{-\nu}^{q \ra q\, g}(k_1,\nu; k_2,\nu_2)\, ,\nonumber
\end{eqnarray}
with $M^{q\,g\ra q\,g}$ given in eq.~(\ref{elasqa}), and with 
collinear factor
\be
{\rm Split}_{-\nu}^{q \ra q\, g}(k_1,\nu; k_2,\nu_2) = g\,
(\lambda^{d_2})_{d_1\bar{c}}\,
{\rm split}_{-\nu}^{q \ra q\, g}(k_1,\nu; k_2,\nu_2)\, ,\label{qgfac}
\ee
with splitting factors \cite{mp},
\begin{eqnarray}
{\rm split}_-^{q \ra q\, g}(k_1,+; k_2,+) &=& \sqrt{2}\, 
{z^{1/2}\over \sqrt{z(1-z)}\langle k_1 k_2\rangle} \nonumber\\
{\rm split}_+^{q \ra q\, g}(k_1,-; k_2,+) &=& \sqrt{2}\, 
{z^{3/2}\over \sqrt{z(1-z)}\langle k_1 k_2\rangle}\, ,\label{qgsplit}
\eea
and with ${\rm split}_{\nu}^{q \ra q\, g}
(k_1,-\nu; k_2,-\nu_2)$ obtained from ${\rm split}_{-\nu}^{q \ra q\, g}
(k_1,\nu; k_2,\nu_2)$ by exchanging $\langle k_1
k_2\rangle$ with $[k_2 k_1]$. 

In the soft quark limit, $k_1 \ra 0$, the vertex $q\, g^* \rightarrow q\, g$
in eq.~(\ref{forwqg}) yields at most a square-root singularity, 
and thus as seen previously it does not yield an infrared divergence.
In the soft gluon limit, $k_2 \ra 0$, eq.~(\ref{forwqg}) reduces to
\begin{eqnarray}
& & \lim_{k_2\ra 0} M^{q\,g\ra q\,g\,g}(-p_a,-\nu; k_1,\nu; 
k_2,\nu_2; p_{b'},\nu_{b'}; -p_b,-\nu_b) = 2\hat s\,  g^2
\nonumber\\ & & \times \left[i f^{d_2c'c} \lambda^c_{d_1\bar{a}}
{\rm soft}(k_1; k_2,\nu_2; p_{b'}) - \left(\lambda^{c'}
\lambda^{d_2}\right)_{d_1\bar{a}} {\rm soft}(-p_a; k_2,\nu_2; k_1) 
\right] \nonumber\\ & & \times
{1\over \hat t}\, \left[ig\, f^{bb'c'} C^{g\,g}_{-\nu_b\nu_{b'}}(-p_b,p_{b'})
\right]\, ,\label{softqg}
\end{eqnarray}
with ${\rm soft}(k_1; k_2; p_{b'})$ and 
${\rm soft}(-p_a; k_2; k_1)$ obtained from eq.~(\ref{nonfac})
by exchanging $k_1 \leftrightarrow k_2$.

In the same fashion as eq.~(\ref{forwqg}), the amplitude for
$\bar q\,g\ra \bar q\,g\,g$ can be analysed \cite{thuile}.

\subsection{Virtual NLL corrections to QCD amplitudes}
\label{sec:1c}

The virtual radiative corrections in NLL approximation can be obtained 
from the one-loop four-parton amplitudes \cite{bk}. As examples,
I shall give here the gluon-gluon and the quark-quark amplitudes.
In NLL approximation it suffices to consider the dispersive part,
which in dimensional regularization and for the gluon-gluon 
amplitude is \cite{flv,dds}
\begin{eqnarray}
\lefteqn{ {\rm Disp}\, M_{1-loop}^{g g \ra g g}(-p_a,-\nu_a; p_{a'},
\nu_{a'}; p_{b'},\nu_{b'}; -p_b,-\nu_b) } \label{seven}\\ &=& 
M_{tree}^{g g \ra g g}(-p_a,-\nu_a; p_{a'},\nu_{a'}; p_{b'},\nu_{b'}; 
-p_b,-\nu_b)\nonumber\\ &\times& g^2\, c_{\Gamma}\,
\left\{ \left({\mu^2\over -t}\right)^{\epsilon} \left[ N_c\left(
-{4\over\epsilon^2} + {2\over\epsilon}\ln{s\over -t} - {64\over 9} 
- {\delta_R\over 3} + \pi^2 \right) - {\beta_0\over\epsilon}
+ {10\over 9}N_f\right] - {\beta_0 \over\epsilon}\right\}\, ,\nonumber
\end{eqnarray}
with $M_{tree}^{g g \ra g g}$ given in eq.~(\ref{elas}),
$N_f$ the number of quark flavors, $c_{\Gamma}$ given in 
eq.~(\ref{cgam}), $\beta_0 = (11N_c-2N_f)/3$, and
\begin{equation}
\delta_R = \left\{ \begin{array}{ll} 1 & \mbox{HV or CDR scheme},\\
0 & \mbox{dimensional reduction scheme}\, , \end{array} \right. \label{cp}
\end{equation}
with (HV) the 't~Hooft-Veltman or (CDR) the conventional dimensional 
regularization schemes \cite{bk}. The last term in eq.~(\ref{seven}) is
the $\overline{\rm MS}$ ultraviolet counterterm.

To LL accuracy, using eq.~(\ref{alph}), 
eq.~(\ref{seven}) reduces to
\begin{equation}
{\rm Disp}\, M_{1-loop}^{g g \ra g g}(-p_a,-\nu_a; p_{a'},
\nu_{a'}; p_{b'},\nu_{b'}; -p_b,-\nu_b) = \alpha(t)\, 
\ln{s\over -t}\, M_{\rm tree}^{g g \ra g g}\, ,\label{ll}
\end{equation}
in agreement with eq.~(\ref{sud}).

The virtual radiative corrections to eq.~(\ref{elasq}) in
NLL approximation are \cite{dds}
\begin{eqnarray}
\lefteqn{{\rm Disp}\, M_{1-loop}^{q\,q\ra q\, q}(-p_a,-\nu_a; p_{a'},
\nu_a; p_{b'},\nu_b; -p_b,-\nu_b) } \nonumber\\ &=& M_{tree}^{q\,q\ra q\, q}
(-p_a,-\nu_a; p_{a'},\nu_a; p_{b'},\nu_b; -p_b,-\nu_b)
\nonumber\\ &\times& g^2 c_{\Gamma}\,\left\{ \left(-{\mu^2\over t}
\right)^{\epsilon} \left[ - {N_c^2-1\over 2N_c} \left({4\over\epsilon^2} + 
{6\over \epsilon}\right) + N_c \left({2\over\epsilon} \ln{s\over -t} 
+ {19\over 9} - {2\delta_R\over 3} + \pi^2 \right) 
\right.\right. \nonumber\\ & & \left.\left. + {\beta_0\over\epsilon}
- {10\over 9} N_f + {1\over N_c} \left(7 + \delta_R
\right) \right] - {\beta_0\over\epsilon}\right\}\, 
.\label{dispq}
\end{eqnarray}
To LL accuracy, eq.~(\ref{dispq}) reduces to 
\begin{equation}
{\rm Disp}\, M_{1-loop}^{q\,q\ra q\, q} = \alpha(t)\, 
\ln{s\over -t}\, M_{tree}^{q\,q\ra q\, q}\, ,\label{llqq}
\end{equation}
which is exactly the same form as eq.~(\ref{ll}),
due to the universality of the LL contribution.

\section{Conclusions}
\label{sec:2}

We have argued that in order to perform a detailed analysis of dijet 
production with a rapidity gap we need an ${\cal O}(\al_s^4)$
calculation including the relevant collinear enhancements
which give structure to the jets. Such a calculation
needs not be exact, but must include the full leading power in 
$\hat{s}/\hat{t}$ of dijet production in the high-energy
limit. The QCD amplitudes must therefore be determined to the
corresponding accuracy. We have shown this in detail for the
scattering amplitudes necessary to compute the full leading power in 
$\hat{s}/\hat{t}$ of dijet production to ${\cal O}(\al_s^3)$.

\appendix
\section{Spinor Algebra in the Multi-Regge kinematics}
\label{sec:appa}

We consider the scattering of two gluons of momenta $p_a$ and $p_b$
into $n+2$ gluons of momenta $p_{i}$, where $i=a',b',1\dots n$.
Using light-cone coordinates $p^{\pm}= p_0\pm p_z$, and
complex transverse coordinates $p_{\perp} = p_x + i p_y$,
the gluon 4-momenta are,
\begin{eqnarray}
p_a &=& \left(p_a^+, 0; 0, 0\right)\, ,\nonumber \\
p_b &=& \left(0, p_b^-; 0, 0\right)\, ,\label{in}\\
p_i &=& \left(|p_{i\perp}| e^{y_i}, |p_{i\perp}| e^{-y_i}; 
|p_{i\perp}|\cos{\phi_i}, |p_{i\perp}|\sin{\phi_i}\right)\, ,\nonumber
\end{eqnarray}
where to the left of the semicolon we have the + and -
components, and to the right the transverse components.
$y$ is the gluon rapidity and $\phi$ is the azimuthal angle between the 
vector $p_{\perp}$ and an arbitrary vector in the transverse plane.
Momentum conservation gives
\begin{eqnarray}
0 &=& \sum p_{i\perp}\, ,\nonumber \\
p_a^+ &=& \sum p_{i}^+ \, ,\label{kin}\\ 
p_b^- &=& \sum p_{i}^- \, .\nonumber
\end{eqnarray}

For each massless momentum $p$ there is a positive and negative helicity 
spinor, $|p+\rangle$ and $|p-\rangle$, so we can consider two types of 
spinor products 
\begin{eqnarray}
        \langle pq\rangle & = & \langle p-|q+\rangle\nonumber\\
        \left[ pq \right] & = & \langle p+|q-\rangle\ .
        \label{spinors}
\end{eqnarray}
Phases are chosen so that $\langle pq\rangle=-\langle qp\rangle$
and $[pq]=-[qp]$.
For the momentum under consideration the spinor products are
\begin{eqnarray}
\langle p_{i} p_{j}\rangle &=& p_{i\perp}\sqrt{p_{j}^+
\over p_{i}^+} - p_{j\perp} \sqrt{p_{i}^+\over p_{j}^+}\, 
,\nonumber\\ \langle p_a p_i\rangle &=& -\sqrt{p_a^+
\over p_i^+}\, p_{i\perp}\, ,\label{spro}\\ \langle p_i p_b\rangle &=&
-\sqrt{p_i^+ p_b^-}\, ,\nonumber\\ \langle p_a p_b\rangle 
&=& - \sqrt{p_a^+ p_b^-} = -\sqrt{s_{ab}}\, ,\nonumber
\end{eqnarray}
where we have used the mass-shell condition 
$|p_{i\perp}|^2 = p_i^+ p_i^-$.  The other type of spinor product 
can be obtained from
\begin{equation}
        [pq]\ =\ \pm\langle qp\rangle^{*}\ ,
\end{equation}
where the $+$ is used if $p$ and $q$ are both ingoing or both 
outgoing, and the $-$ is used if one is ingoing and the other 
outgoing.

In the multi-Regge kinematics, the gluons are strongly ordered in 
rapidity and have comparable transverse momentum:
\begin{equation}
y_{a'} \gg y_1\gg\dots y_{n} \gg y_{b'};\qquad |p_{i\perp}|\simeq|p_{\perp}|\, 
.\label{mreg}
\end{equation}
Then the momentum conservation (\ref{kin}) in the $\pm$ directions reduces to
\begin{eqnarray}
p_a^+ &\simeq& p_{a'}^+\, ,\nonumber\\ 
p_b^- &\simeq& p_{b'}^-\, ,\label{hkin}
\end{eqnarray}
and the Mandelstam invariants become
\begin{eqnarray}
s_{ab} &=& 2 p_a\cdot p_b \simeq p_{a'}^+ p_{b'}^- \nonumber\\ 
s_{ai} &=& -2 p_a\cdot p_i \simeq - p_{a'}^+ p_i^- \label{mrinv}\\ 
s_{bi} &=& -2 p_b\cdot p_i \simeq - p_i^+ p_{b'}^- \nonumber\\ 
s_{ij} &=& 2 p_i\cdot p_j \simeq |p_{i\perp}| |p_{j\perp}| e^{y_i-y_j}
=p_{i}^{+}p_{j}^{-}\qquad (y_{i}\gg y_{j})\,
,\nonumber
\end{eqnarray}
where $i,j=a',b',1\dots n$.  
In this limit the spinor products (\ref{spro}) become
\begin{eqnarray}
\langle p_a p_b\rangle &\simeq& \langle p_{a'} p_b\rangle \simeq
-\sqrt{p_{a'}^+\over p_{b'}^+} |p_{b'\perp}| \nonumber\\
\langle p_a p_{b'}\rangle &\simeq& \langle p_{a'} p_{b'}\rangle =
-\sqrt{p_{a'}^+\over p_{b'}^+}\, p_{b'\perp} \nonumber\\
\langle p_a p_{a'}\rangle &\simeq& - p_{a'\perp} \nonumber\\
\langle p_{b'} p_b\rangle &\simeq& - |p_{b'\perp}| \label{hpro}\\
\langle p_a p_i\rangle &\simeq& \langle p_{a'} p_i\rangle =
-\sqrt{p_{a'}^+\over p_i^+}\, p_{i\perp} \nonumber\\
\langle p_i p_b\rangle &\simeq& -\sqrt{p_i^+\over p_{b'}^+} 
|p_{b'\perp}| \nonumber\\ 
\langle p_i p_{b'}\rangle &\simeq& -\sqrt{p_i^+\over p_{b'}^+} 
p_{b'\perp} \nonumber\\ 
\langle p_i p_{j}\rangle &\simeq& -\sqrt{p_i^+\over p_{j}^+}\,
p_{j\perp} \qquad (y_{i}\gg y_{j})\ .\nonumber
\end{eqnarray}

\end{document}